# A PARAMETRIC STUDY ON WINDOW-TO-FLOOR RATIO OF DOUBLE WINDOW GLAZING AND ITS SHADOWING USING DYNAMIC SIMULATION

**Ana Rita Amaral[a*], Eugénio Rodrigues[a,b],
Adélio Rodrigues Gaspar[a], Álvaro Gomes[b,c]**

a) ADAI, LAETA & Department of Mechanical Engineering, Faculty of Sciences and Technology, University of Coimbra, Rua Luís Reis Santos, Pólo II, 3030-788 Coimbra
b) INESC Coimbra – Institute for Systems Engineering and Computers at Coimbra, Rua Antero de Quental 199, 3000-033 Coimbra
c) Department of Electrical and Computer Engineering, Faculty of Sciences and Technology, University of Coimbra, Pólo II, 3030-290 Coimbra

* e-mail: aritamaral@hotmail.com

**Key words:** thermal comfort, dynamic simulation, window performance, energy efficiency, window-to-floor ratio

*Abstract. When incorrectly designed, windows can be responsible for unnecessary energy consumption in a building. This may result from its dimensions, orientation and shadowing. In a moderate climate like the Portuguese, and considering an annual thermal comfort assessment of a space, if windows are under-dimensioned or over-shadowed, they can contribute to the increase of heating needs. However, when over-dimensioned or under-shadowed, they contribute to the increase of cooling requirements. Therefore, it is important to find the optimum design that balances orientation, dimension and shadowing, contributing to minimize both the heating and cooling needs.*

*This study presents a parametric analysis of a double glazing window in its orientation and dimension, located in the Portuguese city of Coimbra. For each window orientation and dimension, the optimum overhang depth is determined. The objective is to minimize degree-hours of thermal discomfort. Results show that overhangs are mainly a corrective mechanism to over-dimensioned openings, thus allowing that building practitioners may choose a wider range of windows dimensions.*

## 1 INTRODUCTION

Due to materials characteristics, openings are the most fragile elements in the building's envelope in terms of thermal performance; it is known that glazing areas are still the elements presenting the highest values of heat transfer coefficient (U-value). In this sense, if an inappropriate window area may increase negatively heat gains or losses, an incorrect shadowing can contribute to an inefficient use of solar radiation, which can affect the thermal comfort of its occupants. So, providing useful data to designers in an early architectural design phase may significantly contribute to the adoption of more energy efficient solutions.

Overhangs and fins represent non mechanic shading devices whose implementation can be considered as low cost when compared to automated shadowing [1]. However, when designing solar shadings, thermal comfort and visual comfort may conflict [2].

In what concerns to openings issues, the advances of windows technologies are predominant in literature; glazing type, fenestrations products and materials, spacers, frames, air or gas gap between glass layers, and so on, are seen as the key elements that may influence the global performance of a window [3]. However, by their benefits in the reduction of potential solar heat gains and in implementation costs, a depth study is deserved to shadowing. Some tools have been developed to optimize solutions of glazing areas according to illumination, glare, solar radiation, or visual comfort, as Menghani et al and Andersen et al propose [4,5]; however, shading devices seem to be focused more by natural lighting concerns than by thermal comfort or energy efficiency. When focusing in these issues, it is seen the use of dynamic simulation tools to predict gains and losses through windows [6], and also to assess the effect of overhangs in energy savings [1,7]. Nevertheless, these studies have in common the use of a predefined window dimension and the simple possibility of having an overhang, a side fin, both or none, leaving out of the analysis the influence of possible windows or overhangs dimensions.

The design of an adequate dimension or position for shading devices, such as overhangs, is still a challenge for architects to adequate those elements to windows performance needs. So, currently, overhangs are designed for aesthetics issues rather than by a conscious solar protection method. In this sense it is important to achieve an easy and user-friendly means to obtain the optimum overhang for each opening. The present study analyzes overhang's contribution to the thermal performance of a reference room with different window dimensions and orientation. The main purpose is to assist practitioners, namely architects, to take conscious decisions in an early phase of design process, by providing data in terms of windows and overhangs best dimensions.

## 2   METHODOLOGY

The present study considers a thermal space as a reference room with fixed dimensions, an exterior façade with a window, and predefined constructive systems. Figure 1 presents its geometry, as well as the window position in the widest wall and the overhang to be analyzed. Window dimensions are variable; starting with a width of 0.01m and a height of 2.00m, the opening is resized until it reaches 7.00m wide (w), corresponding to a WFR variance between 0 and 0,614. Withal, window orientation is analyzed around 360º in a two-degree step, starting at 0º (North) and turning clockwise. For each window size step (w) the optimum overhang depth (d) is determined and has the same width as the window.

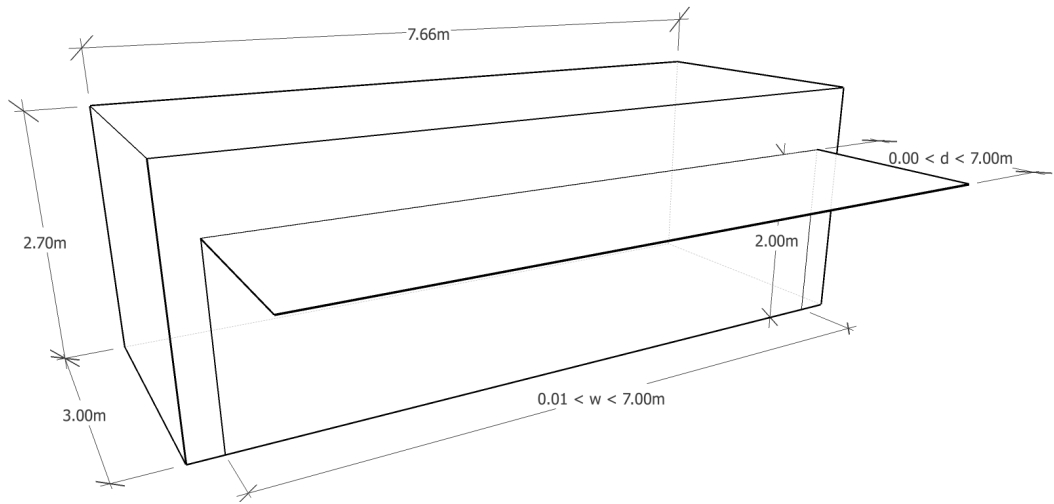

Figure 1: Thermal space geometric specifications for the parametric study

Table 1 presents the envelope constructive systems for the reference room as well as double glazing window, defined according to the reference U-values of the Portuguese regulation [8], knowing that Coimbra is included in the climatic zone I2 for winter and V2 for summer.

Table 1: Reference U-values and materials physical properties

| Element | U-value | Type | SHGC | VT |
|---|---|---|---|---|
| Ext. wall | 0.43 | Double brick wall | | |
| Floor | 0.45 | Ground floor | | |
| Roof | 0.37 | Flat roof | | |
| Window | 2.60 | Double glass | 0.63 | 0.56 |
| SHGC: solar heat gain coefficient; VT: visible transmittance | | | | |

To carry out this study, it was considered a parametric approach, where two algorithms, EPSAP and FPOP, were specifically adjusted [10–14]. EPSAP (Evolutionary Program for the Space Allocation Problem) consists in a hybrid evolutionary strategy approach enhanced with local search technique to allocate rooms on a floor plan [10–12]. FPOP (Floor plan Performance Optimization Program) consists in a sequential variable optimization procedure, where different building geometry variables, such as the openings orientation, position and size, are changed with the aim of minimizing the thermal penalties. These are obtained by the determination of the total degree-hours of discomfort (TDH) [13,14]. It was also obtained the heating degree-hours (HDH) and the cooling degree-hours (CDH) by the difference between the operative indoor temperature in the room and adaptive thermal comfort limits for naturally ventilated spaces, according to the European Standard 15251:2007. That operative temperature was determined using dynamic simulation program EnergyPlus 8.1.0, as well as the thermal performance of each solution.

In order to achieve a global optimum overhang size, it was considered that other factors that may interfere in the results by a random number chosen for studying are valueless – for instance occupation schedules or internal gains from hypothetic equipment and lighting. In this sense, the room was considered to have no occupation and no internal gains. However, it was considered an infiltration rate of a steady 0,4 Air Changes per Hour (ACH), according to Portuguese regulation [9]. The openings were considered close all time, thus not having any additional natural ventilation, since the room was not occupied. It was considered the weather data for Coimbra, Portugal, retrieved from the US Department of Energy website. This data, location information and physical properties of constructive systems materials are stored in a database in FPOP algorithm.

The room requirements were introduced in the EPSAP algorithm, which generated 180 reference rooms. After the reference rooms have been generated, the FPOP algorithm has transformed the size of the existing window until it reached 7.00m wide. Such approach was previously used to study and compare the thermal performance of three window types and to determine the optimum opening dimension for all orientations [15]. However, in this study, the focus is given in a double glazing window and, for each step of its sizing, the optimum overhang depth was also determined.

## 3 RESULTS AND DISCUSSION

Through the described approach, it was possible to achieve an annual thermal assessment for different orientations, as well as a precise range of overhang depths according to windows size and orientation. Figure 2 shows the optimum performance points for the 360º possible orientations. The left-axis represents the space relative thermal performance from an area without any opening to the best opening size of all orientations. The top and bottom axis represent the WFR and Window-to-Wall Ratio (WWR), respectively. The right-axis in red is the ratio between CDH and TDH, and the green-axis represents the relative overhang depth. Finally, black line shows all the optimum values of WFR for all orientations and, according to the correspondent value in green line, it is possible to obtain the respective overhang dimension.

As shown in Figure 2, an unexpected behavior is evident in North quadrant, since it has one of the highest values of WFR, comparing to the other orientations. From a detailed analysis of the heat gains and losses, it is concluded that this is due to window sky diffuse radiation gain being higher than the additional losses (relatively to opaque wall surface) from heat transfer to the exterior by the window surface. In this case, the overhang does not contribute to the thermal performance as the optimum window size is near to 0.4 WFR and the shading mechanism is only necessary after 0.55 WFR. Similarly, in Northeast and Northwest orientations the overhang is also unnecessary. In both cases, if the window is slightly bigger than 0.23 and 0.18 WFR (Northeast and Northwest respectively), the overhang may contribute to reduce CDH.

However, the behavior is different for orientations ranging from West to East passing through South. In these orientations, the overhangs reduce CDH (red line) as their depth increase (green line). This allows to optimum WFR of openings to be higher. For instance, the South orientation the optimum size without overhang is near 0.18 WFR however and with shading mechanism is above 0.35 WFR. From all orientations, the West and East are the ones that require the larger overhangs (0.30 and 0.25 relative depth respectively).

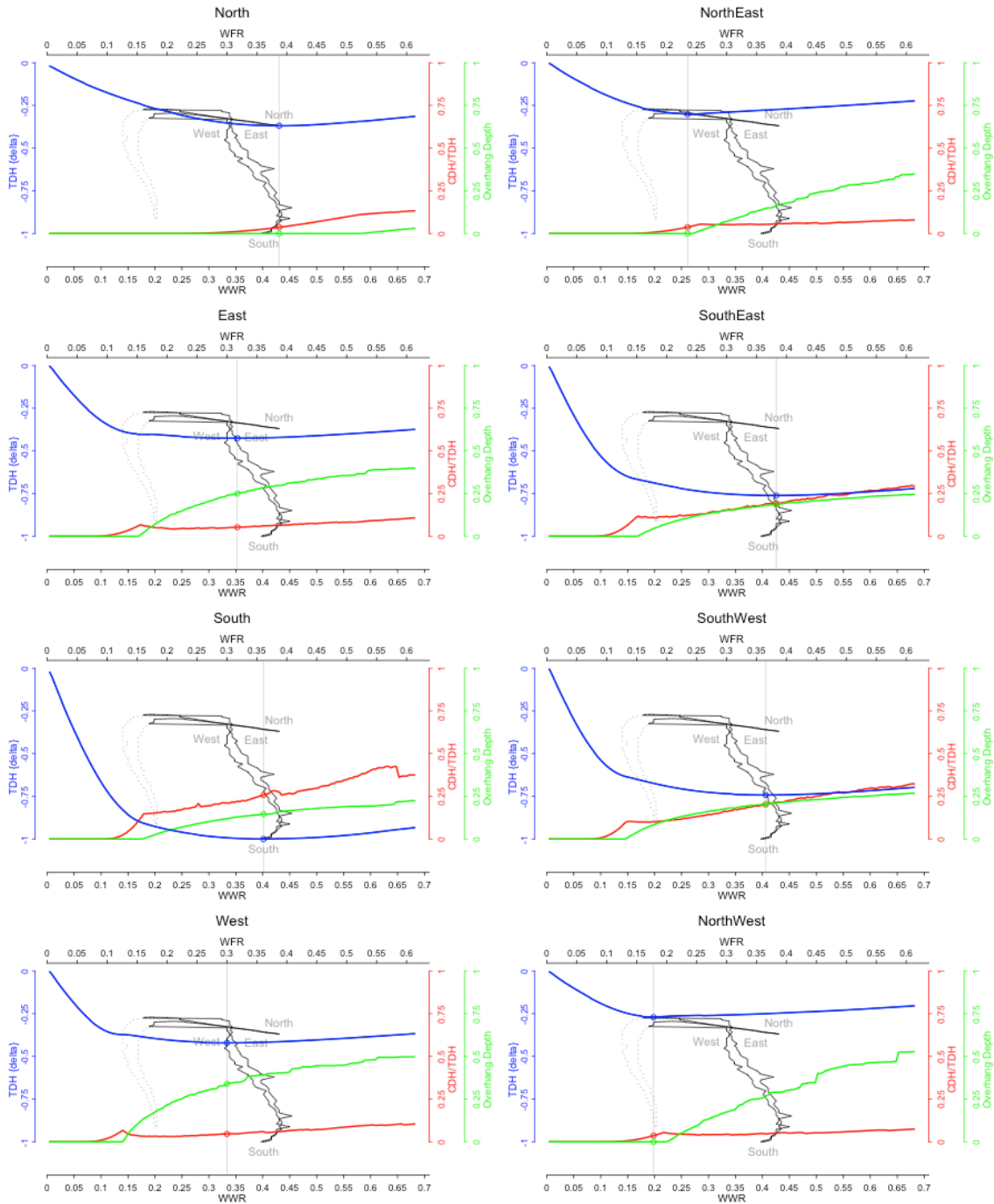

Figure 2: Annual thermal assessments of the window size and respective overhang depth. Blue line represents the relative space thermal comfort performance (TDH), the red line in the ratio of CDH per TDH, and green line shows overhang relative depth. Black line represents the set of optimum window size for all orientations. Dotted black line represents the optimum window size for all orientations without the use of an overhang.

As the main contribution of the overhangs to the thermal performance is the reduction of the cooling needs, the overall improvement is not significant in comparison to a well-dimensioned window from thermal comfort perspective (when comparing the dotted black line and the continuous black line in graphics). However, shading mechanisms allow the building designer to explore larger window areas. This is especially evident in East and West orientations, where the thermal performance (strong blue line) is almost horizontal. However, the study should be extended to include other criteria, such as visual comfort and lighting requirement.

## 4 CONCLUSIONS

Even though it was not considered other factors that may influence a global performance, such as equipment, lighting or even occupancy schedules, the overhangs contributed to the improvement of the thermal performance in the reference room for the West, South, and East quadrants by reducing the cooling needs. However, from thermal comfort criteria only, the resulting improvement is not significant when compared to optimum window dimensions. Thus, it is possible to conclude that the use of overhangs allows building design practitioners to have larger glazing areas or to use them as corrective measures to over-dimensioned openings.

In a global way, this methodology allows a simple evaluation of building systems, namely openings, providing data that can help architects to obtain an optimal window size with fewer degree-hours of thermal discomfort for all orientations, according to orientation, location and buildings physical properties, and especially to obtain a precise depth for the studied shading device. This is particularly important in an early phase of the design process, as it aims to help the adequate incorporation of those shading devices in the building design. As this approach combines a simulation tool, it is possible to extrapolate this study for all existing climate data, being helpful to the overall designer's community. Moreover, this accuracy in designing adequate and integrated shadings for each window, allows a further possibility of taking advantage of openings in their other benefits, such as the natural light available or architectural aesthetics considerations.

**ACKNOWLEDGEMENTS**

This work is framed under the *Energy for Sustainability Initiative* of the University of Coimbra (UC) and has been supported by the project Automatic Generation of Architectural Floor Plans with Energy Optimization - GerAPlanO - QREN 38922 Project (CENTRO-07-0402-FEDER-038922).